\documentclass[aps,prl,superscriptaddress,twocolumn,floatfix]{revtex4}
\usepackage{amsmath}
\usepackage{graphicx}

\begin{document}

\flushbottom

\title{Electrons in Dry DNA from Density Functional Calculations}

\author{ Emilio Artacho }
\affiliation{ Department of Earth Sciences, Downing Street, 
              University of Cambridge,
              Cambridge CB2 3EQ, UK }

\author{ Maider Machado }
\affiliation{ Dep. de F\'{\i}sica de Materiales,
              Universidad del Pa\'{\i}s Vasco,
              20080 Donostia, Spain }

\author{ Daniel S\'anchez-Portal }
\affiliation{ Centro Mixto CSIC-UPV/EHU and Donostia Intl. Physics Center,
              Po. Manuel de Lardizabal 4, 
              20080 Donostia, Spain }

\author{ Pablo Ordej\'on }
\affiliation{ Institut de Ci\`encia de Materials de Barcelona, CSIC,
              08193 Bellaterra, Barcelona, Spain }

\author{ Jos\'e M. Soler }
\affiliation{ Dep.\ de F\'{\i}sica de la Materia Condensada, C-III,
              Universidad Aut\'{o}noma de Madrid,
              E-28049 Madrid, Spain }

\date{19 September 2002}

\begin{abstract}
  The electronic structure of an infinite poly-guanine -
poly-cytosine DNA molecule in its dry A-helix structure is studied
by means of density-functional calculations.
  An extensive study of 30 nucleic base pairs is performed to
validate the method.
  The electronic energy bands of DNA close to the Fermi level are
then analyzed in order to clarify the electron transport
properties in this particularly simple DNA realization, probably
the best suited candidate for conduction.
  The energy scale found for the relevant band widths, as compared
with the energy fluctuations of vibrational or genetic-sequence
origin, makes highly implausible the coherent transport of
electrons in this system.
  The possibility of diffusive transport with sub-nanometer
mean free paths is, however, still open.
  Information for model Hamiltonians for conduction is provided.
\end{abstract}


\maketitle

\section{Introduction}

  The idea of using DNA chains as molecular wires in
nano-devices has been proposed and is being explored in the
nano-science community \cite{Mirkin96,Alivisatos96,Braun98}, given
the extremely high selectivity of a DNA strand to pair with its
complementary partner, which facilitates the design of complex
circuitry at the nanometer scale.
  Even though some researchers have chosen to coat \cite{Braun98} or
modify \cite{Rakitin01} the DNA wires in order to improve its
conduction properties, the original idea was to use pure DNA,
assuming that electric current could flow along the wires.
  This assumption has proven to be less trivial than expected,
however, and has originated an important controversy.

  Direct measurements show from low resistance DC conduction
\cite{Fink99} (even superconduction \cite{Kasumov01}), to no
conduction at all \cite{Braun98,dePablo00,Gomez02}, including
results with non-ohmic characteristics \cite{Porath00,Yoo01} or
appreciable ac conductance \cite{Gruner00}.
  The transport has been proposed to be band-structure-like
\cite{Porath00}, by tunneling \cite{Beratan97}, solitonic
\cite{BenJacob98}, polaronic \cite{Schuster99,Conwell00,Yoo01}, or
facilitated by fluctuations \cite{Ladik99,Bruinsma00}.
  This rich scenario is not surprising when considering the
comparably small amount of clean-cut information available (mostly
theoretical) about basic aspects of the electronic structure of
dry DNA \cite{Ladik71,Ladik83,Pablo97,Yang98,Ye00}, (in all the
experiments mentioned above the samples were dried in one way or
another).
  In addition to the fundamental complexity of the system
(large system size, weak interactions, structure floppiness),
there is a substantial lack of detailed knowledge about the
environment the DNA molecules are left with after the drying
process.
  Most likely, residual water molecules will stay adhered to the
chain, together with the cations counterbalancing the negative
charge of the molecule, all quite uncontrolled.

  A theoretical simulation benefits from absolute control at
the atomic scale.
  In this particular case, the system can be prepared in the best
possible way for conduction.
  Previous simulations \cite{Ladik71,Ladik83,Pablo97,Yang98,Ye00} have
offered very insightful information for related problems, in spite
of being based on relatively drastic approximations in the
electronic structure, due to the system size.
  They were mostly aimed at the description of DNA properties in
live conditions, as were the recent first-principles studies in
wet DNA \cite{Landman01,Zhang02,Parrinello02}.

  Recent developments in linear-scaling
first-principles \cite{Soler02,Bowler02,Scuseria00} methods based
on density-functional theory \cite{Payne92} (DFT) allow the study
of an infinite DNA double helix from first principles.
  DFT functionals based on the Generalized Gradient Approximation (GGA)
have shown to provide very satisfactory accuracy in molecules akin
to DNA \cite{Sponer96,Hutter96}.
  In this work, an extensive study is presented on a set of 30
nucleic base pairs that validates the theoretical method.
  Then, an infinite DNA chain of trivial sequence is studied: all
guanines (G) on one strand, all cytosines (C) on the other, in the
Watson-Crick pairing arrangement.
  We simulate it completely dry and in vacuo, and in an A-helix
structure, the most relevant one for these conditions.
  The electronic structure is analyzed after relaxing the
structure with the first-principles forces.
  Genetic sequence change effects are explored on a similar system
where one guanine-cytosine pair is swapped every eleven base
pairs, then relaxed and analyzed.
  Preliminary results of this study were published in a joint
experiment-theory paper \cite{dePablo00}.

\section{Method}

  The first-principles DFT calculations reported here were
performed using the {\sc Siesta} method
\cite{Soler02,Pablo96,Daniel97}, which has already been applied to
a large variety of systems \cite{websiesta}, including
biomolecules \cite{dePablo00,Marivi00}.
  It is a fully linear-scaling method, both in the building of the
DFT Hamiltonian and in its solution.
  GGA ~\cite{PBE} was used for electron exchange and
correlation.
  Core electrons were replaced by norm-conserving
pseudopotentials \cite{TroullierMartins91} in their fully
non-local formulation~\cite{KB}.
  A uniform mesh with a planewave cutoff of 150 Ry was used
for integrations in real space \cite{Soler02}.
  The basis sets used are made of numerical atomic orbitals of
finite range and are described below.

  The unit cell of the A-helix studied here includes eleven
poly(G)-poly(C) (pGpC) base-pairs with a total of 715 atoms.
  Periodic boundary conditions are used to simulate infinitely
long DNA chains well separated from each other, with negligible
mutual interactions.
  Only the $\Gamma$ $k$-point of the 715-atom cell was included in the
calculations, which corresponds to 11 $k$-points, if the helical
symmetry is considered.
  The initial geometry was that of an experimental low-resolution
A-helix structure for the pGpC sequence
\cite{Landoldt-Bornstein},
  The geometry was relaxed by means of ab initio linear-scaling DFT,
as described elsewhere \cite{ADNA-PRB}.
  The final structure had still well defined A-helix
characteristics \cite{ADNA-PRB,Calladine97}.
  A second DNA realization (swapped hereafter) was built by
swapping one single G-C pair into C-G out of the eleven in the
cell, followed by another full relaxation.

  The electronic structure is analyzed in terms of the Kohn-Sham
eigenvalues and eigenfunctions, based on Janak's theorem
\cite{Janak78}.
  They are obtained by a single cube-scaling diagonalization,
using the final geometry and electron density obtained
with the linear-scaling method.
  The known band-gap problem of DFT is not important here, since
the particular value of the band gap is of no relevance for the
conclusions of this study as long as there is a gap, which is the
case.
  The band widths have been shown to be well described by this
method, at least up to the kind of accuracy needed in this work.
  A much more costly calculation including explicit electronic
correlation would not be justified.
  Mott-like correlations, that could seem to be relevant
for the narrow band-widths characteristic of the problem (see
below), are not important given the small density of carriers
(electrons or holes, see discussion below) expected in the system.
  The effect of disorder and vibrations is explicitly considered
below.
  The next section is devoted to a study of nucleic base pairs
where the focus is mainly on the hydrogen bonds, which might in
principle be worst described by our methodology.
  It will be followed by the results for A-DNA.

\section{Study on base pairs}

  A systematic study of 30 nucleic base pairs has been performed
with the methodology presented above, in order to assess the
reliability of the approximations, but also obtaining new results.
  For this study we have used a standard doble-$\zeta$ polarized
(DZP) basis set, namely, a double (split valence) basis for each
valence orbital plus polarization functions in all the atoms.
  The cutoff radii for the atomic orbitals of each element were
obtained for an energy shift \cite{Emilio99} of 50 meV.

  A significant range of configurations of the four bases
guanine, cytosine, adenine and thymine (G, C, A, T) are
considered, the same as those studied by $\rm{\check{S}poner}$
{\em et al.}~\cite{Sponer96} in their MP2 study.
  The Watson-Crick configurations are designated WC, and the
Hoogsteen, reversed Hoogsteen and reversed Watson-Crick appear as
H, RH, RWC respectively.
  Other configurations are distinguished simply with numbers, eg.
AA1, AA2, etc. following the nomenclature of Hobza and
Sandorfy~\cite{Hobza}.
  The structures of the bases and base-pairs studied in this work
can be found in Figures 1 and 2 of Ref.~\onlinecite{Sponer96}. For
the numbering of the atoms we followed Ref.~\onlinecite{Sponer2}.
  In the following our results are compared with those of other
methods.
  The direct comparison with experimental geometries has been
shown to be misleading because of the important charge
rearrangements that happen in the crystal phases used in
experiments \cite{Celia00}.

  The interaction energy $E_{int}$ in the following is defined as
the energy of the base-pair minus the energy of each base with the
same geometry it has in the pair.
  The total stabilization energy $E_t$ is defined as
the difference between the energy of the pair and that of each
base in its isolated optimal geometry.
  The difference between both is thus the deformation energy, {\em
i.e.}, the increase in intramolecular energy due to the geometry
change when the base-pair is formed.

  In this study all the energies have been corrected for basis set
superposition error (BSSE) using the standard Boys-Bernardi
counterpoise correction~\cite{Boys}.
  The correction found for $E_{int}$ is used for $E_t$ as well,
given the fact that the relaxation of the isolated base molecules
has been found to change the BSSE correction by less than 10\%.
  No BSSE correction was included in the forces.
  The effect of this approximation was gauged for the AA1
base-pair, for which BSSE corrected and uncorrected forces were
calculated for independent relaxations, showing a deviation of
0.02 \AA\ in the final acceptor-hydrogen distances.

\begin{table}
\begin{center}
\caption[]{Base-pair Interaction Energies ($E_{int}$, in kcal/mol)
at HF/6-31G** geometries.} \label{eint-hf}
\begin{ruledtabular}
\begin{tabular}{crrr}
Pair & MP2\tablenote{From ref.~\onlinecite{Sponer96}}
     & SIESTA  & Deviation (\%) \\
\tableline
GCWC  & -25.8 & -26.8 & -3.9 \\
GG1   & -24.7 & -25.1 & -1.6 \\
GCNEW & -22.2 & -21.7 &  2.2 \\
CC    & -18.8 & -17.5 &  6.9 \\
GG3   & -17.8 & -16.6 &  6.7 \\
GA1   & -15.2 & -15.5 & -2.0 \\
GT1   & -15.1 & -15.0 &  0.7 \\
GT2   & -14.7 & -14.5 &  1.4 \\
AC1   & -14.3 & -14.0 &  2.1 \\
GC1   & -14.3 & -14.7 & -2.8 \\
AC2   & -14.1 & -14.7 & -4.2 \\
GA3   & -13.8 & -13.8 &  0.0 \\
TAH   & -13.3 & -13.7 & -3.0 \\
TARH  & -13.2 & -13.6 & -3.0 \\
TAWC  & -12.4 & -12.3 &  0.8 \\
TARWC & -12.4 & -12.3 &  0.8 \\
AA1   & -11.5 & -11.7 & -1.7 \\
GA4   & -11.4 & -11.7 & -2.6 \\
TC2   & -11.6 & -10.8 &  7.5 \\
TC1   & -11.4 & -10.6 &  7.0 \\
AA2   & -11.0 & -11.4 & -3.6 \\
TT2   & -10.6 & -9.9  &  6.6 \\
TT1   & -10.6 & -10.1 &  4.7 \\
TT3   & -10.6 & -10.2 &  3.8 \\
GA2   & -10.3 & -10.6 & -2.9 \\
GG4   & -10.0 & -7.4  & 26.0 \\
AA3   & -9.8  & -9.8  &  0.0 \\
2aminoAT & -15.1 & -15.2 & -0.7
\end{tabular}
\end{ruledtabular}
\end{center}
\end{table}

  Our results for $E_{int}$ as compared with the MP2 results of
$\rm{\check{S}poner}$ and coworkers\cite{Sponer96} are presented
in table~\ref{eint-hf}, as calculated for the same geometries
obtained at the Hartree-Fock (HF) 6-31G** level \cite{Sponer96}.
  For all the base-pairs except GG4, the agreement is quite
satisfactory, with differences smaller than 8\% and much less in
most cases.
  GG4 is an exception to the general trend with a relative
difference of 26\% that is reduced to 16\% by extending the basis
range to the longer radii given by an energy shift of 10 meV.
  A similar correction for the other base pairs gives much
smaller variations.
  The standard deviation of our results compared to the MP2 values
is of 0.73 kcal/mol, the same value found by $\rm{\check{S}poner}$
{\em et al.}\cite{Sponer96}using the B3LYP functional \cite{b3lyp}
on the same geometry.
  The standard deviation between both sets of DFT results (our PBE
and their B3LYP) is of 0.85 kcal/mol.

  Concerning geometries, the only pair relaxed with MP2 is
cytosine-cytosine \cite{Sponer96}.
  Table~\ref{cytocyto} compares the length of the hydrogen bond
and both $E_{int}$ and $E_t$ for various methods, showing again a
very satisfactory agreement.

\begin{table}
\begin{center}
\caption{Cytosine-cytosine base-pair, distances in \AA\ and
energies in kcal/mol.} \label{cytocyto}
\begin{ruledtabular}
\begin{tabular}{lccccc}
 & HF\tablenote{Energies and geometry with
   HF 6-31G** \cite{Sponer96}}
 & MP2/HF\tablenote{MP2 energies on HF
   geometry \cite{Sponer96}}
 & MP2\tablenote{MP2 energies and geometry \cite{Sponer96}}
 & DFT\tablenote{B3LYP energies and geometry
   \cite{Sponer96}}
 & This work \\
\hline \vspace{-0.3truecm}
\\
d(N4(H)-N3) & 3.050  & 3.050 & 2.980 & 2.900 & 2.872 \\
\vspace{-0.2truecm}
\\
$E_{int}$  &  -17.3 & -18.8 & -20.5 & -20.4  & -21.1 \\
$E_{t}$    &     -- & -17.5 & -18.7 & -18.1  & -18.5 \\
\end{tabular}
\end{ruledtabular}
\end{center}
\end{table}

  The results shown in this section validate the approximations
used in our method for the system under study, including the
sensitive hydrogen bonds.
  Even if the piling up of bases was not investigated, other
studies show \cite{Hutter96} that GGA functionals can describe
them with the required accuracy when connected by the
sugar-phosphate chains.

\begin{table}
\begin{center}
\caption[ ]{Hydrogen-bond distances (in \AA) for the
guanine-cytosine pair. DZP and DZ(P) stand for two different basis
sets used in this work.}
\begin{ruledtabular}
\begin{tabular}{lccccc}
 & B3LYP\tablenote{J. Sponer {\it et al.} ref. \cite{Sponer96}}
 & BP86\tablenote{C. F. Guerra {\it et al.} ref. \cite{Celia00}}
 & VWN-BP\tablenote{R. Santamar\'{\i}a and A. V\'azquez,
   ref.~\cite{Santamaria}}
 & DZP
 & DZ(P) \\
\hline
N2 -- O2 &  2.93 &  2.87  &  2.930 &  2.872 &  2.892 \\
N2 -- H  &   -   &   -    &  1.035 &  1.035 &  1.023 \\
N1 -- N3 &  2.92 &  2.88  &  2.923 &  2.913 &  2.822 \\
N1 -- H  &   -   &   -    &  1.051 &  1.056 &  1.043 \\
O6 -- N4 &  2.78 &  2.73  &  2.785 &  2.770 &  2.715 \\
H  -- N4 &   -   &   -    &  1.055 &  1.057 &  1.054 \\
\end{tabular}
\end{ruledtabular}
\label{gc}
\end{center}
\end{table}

  The whole set of base pairs as well as the isolated bases have
been relaxed with our method.
  The results for geometry and interaction energies can be found
in \cite{Maider00}.
  We extract in table~\ref{gc} the results for the
guanine-cytosine pair, of relevance for the DNA calculations in
the next section.
  The results of several authors
\cite{Sponer96,Celia00,Santamaria} are compared, which include
different GGA functionals, namely B3LYP \cite{b3lyp}, BP86
\cite{p86,b86}, and VWN-BP \cite{p86,b88,vwn}.
  The results of the DZP basis of this study are also compared
with those of the basis set used for DNA below, DZ(P), a
double-$\zeta$ basis \cite{Emilio99,radii}, which was polarized
for phosphorous and for the atoms involved in the hydrogen bonds.
  The Kohn-Sham eigenvalues around the Fermi level for an isolated
GC pair were checked for DZ(P), which
compare well with the corresponding values
for the DZP basis.
  The convergence of the spatial range of the basis orbitals,
important for band widths in the DNA molecule, was also tested
\cite{radii}.

\section{Results for A-DNA}

\begin{figure}[htbp]
\begin{center}
\includegraphics[clip, bb= 120 85 445 780,width=\columnwidth] {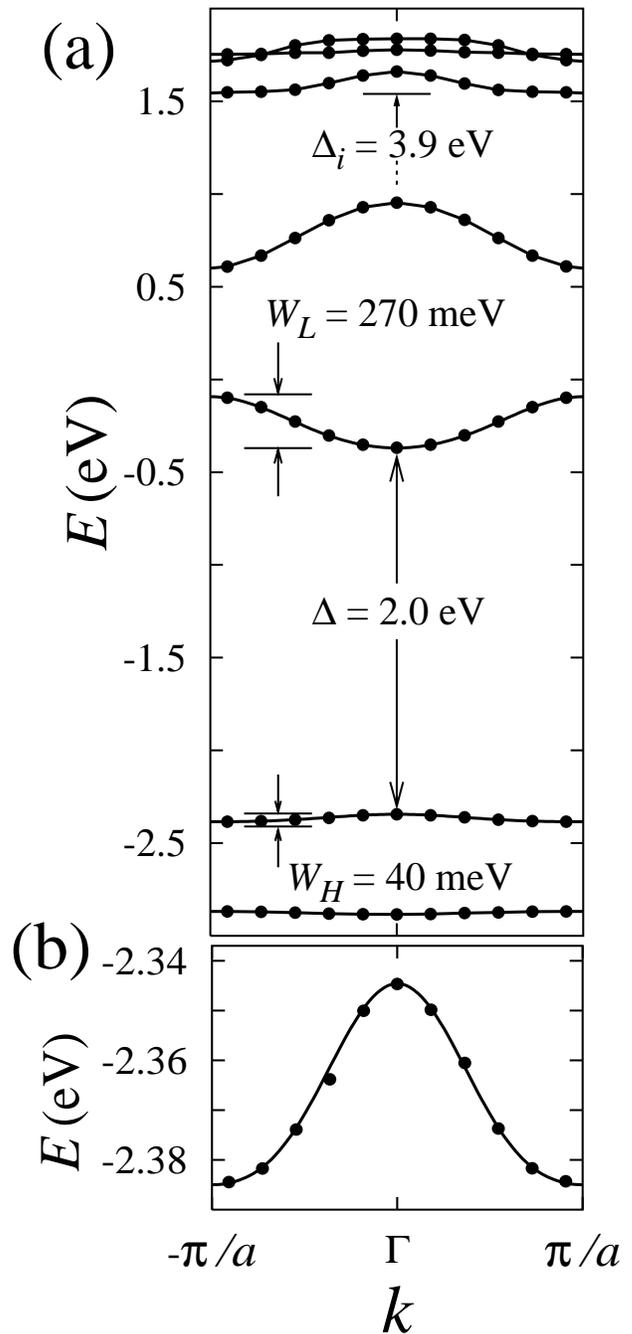}
\caption[] {(a) Kohn-Sham energy bands (points) close to the Fermi
level as functions of the $k$ values corresponding to the helical
symmetry.
  $\Delta$ indicates the HOMO-LUMO gap, $\Delta_i$ the intra-guanine
gap, $W_H$ the band width of the highest occupied band, and $W_L$
the band width of the lowest unoccupied band.
  (b) Kohn-Sham highest occupied band (points), and band of a 1D
tight-binding model with one orbital per unit cell (line), with
$t=10.1$ meV, and $t'=1.5$ meV as first and second nearest
neighbor interactions, respectively.} \label{bands}
\end{center}
\end{figure}

  Figure~\ref{bands} shows the bands of pGpC close to the Fermi level.
  As expected for a well saturated molecule, the Fermi level is in a
clear band gap, which is of 2.0 eV in this case (this number is to
be taken with caution since usual DFT functionals tend to
underestimate the band gaps of insulators by a substantial
amount).
  Electronic transport will thus be mediated by carriers in either
the top of the valence band (holes, or radical cations in
biochemical language) or in the bottom of the conduction band
(electrons or radical anions), introduced by doping.

  In the absence of disorder, the electronic structure close to
the Fermi level shows well defined bands with eleven states per
unit cell, one per base pair in the unit cell.
  The top-most valence band has a very small bandwidth of 40 meV
(see figure~\ref{bands}), the energy separation with the next band
below being ten times larger.
  This first band is associated to the $\pi$-like highest
occupied molecular orbital (HOMO) of the guanines.
  The continuous line in the figure is the tight-binding band of a
one-dimensional system with one orbital per unit cell.
  The excellent agreement of this extremely simple model with the
ab initio points [see figure~\ref{bands} (b)] demonstrates that
each base pair contributes to this band with one single orbital
that interacts negligibly with other orbitals in the pair.
  This allows the modelling of the basic physics of transport in
this molecule with one single orbital per base pair with an
electron hopping interaction of 10 meV.
  Other terms for a model Hamiltonian are quantitatively
justified below.

\begin{figure}
\begin{center}
\includegraphics[width=\columnwidth] {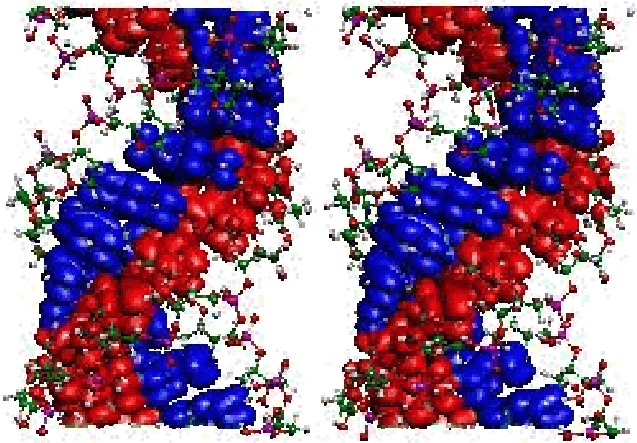}
\caption[]{(Color stereogram) Surfaces of constant density (1.5
$10^5$ $e^-$/\AA$^3$) for the states corresponding to the lowest
unoccupied band (red), and highest occupied band (blue) of the
ordered pGpC structure.  The graphs were produced with Molekel
\cite{molekel}. } \label{homo-lumo}
\end{center}
\end{figure}

  Figure~\ref{homo-lumo} displays the charge density of the
states associated to this band, which appears almost exclusively
on the guanines, with weight neither in the backbones nor in the
cytosines.
  The lowest conduction band has a width of 270 meV, similar to
the width of the spectroscopic feature observed by Porath {\it et
al.} \cite{Porath00}.
  It is separated from the next band by 0.7 eV.
  Similarly to the HOMO situation, this band is made of the lowest
unoccupied molecular orbital (LUMO) of the cytosines, as shown in
figure~\ref{homo-lumo}, although the overlap in this case is more
pronounced, as corresponds to an eight times larger bandwidth.

  Note that any matrix element between HOMO and LUMO states is
very much depressed by the very small spatial overlap between
them.
  Optical absorption will thus be very weak for photon energies
corresponding to the HOMO-LUMO gap (adequately corrected including
correlation and excitonic effects).
  The unoccupied band above the LUMO band remains of a marked
cytosine character.
  The third unoccupied band 3.9 eV above the HOMO band (see
figure~\ref{bands}) is the first one with substantial weight on
the guanines, and thus with an appreciable oscillator strength for
absorption.
  The comparison of these results with those obtained for wet
conditions \cite{Parrinello02} shows that the LUMO is quite
sensitive to the environment, moving from the cytosines to the
cations when in presence of Na$^+$ and water.

\begin{figure}
\begin{center}
\includegraphics[clip, bb= 60 90 510 610,width=\columnwidth] {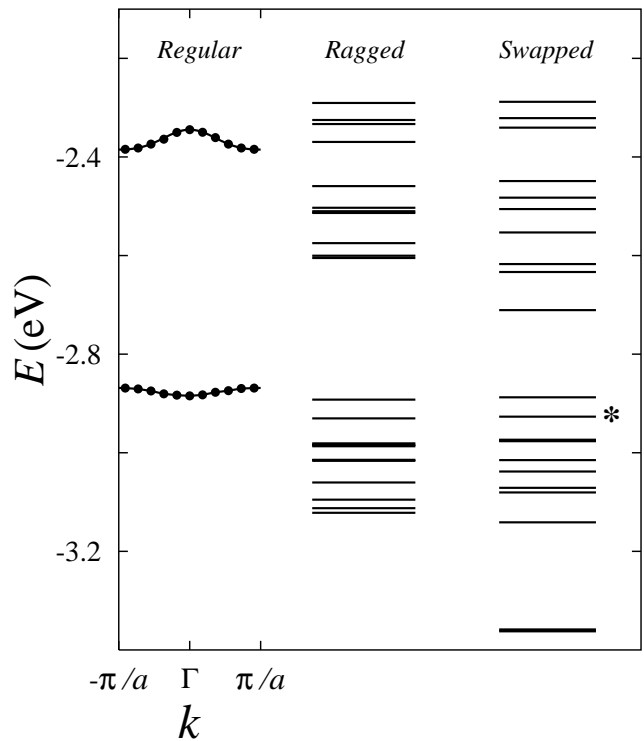}
\caption[]{Highest occupied electronic levels for the regular
pGpC, the ragged, and the swapped structures. The asterisk
indicates a localized state with most of its weight on the HOMO of
the swapped G.} \label{bars}
\end{center}
\end{figure}

\begin{figure}
\begin{center}
\includegraphics[width=\columnwidth] {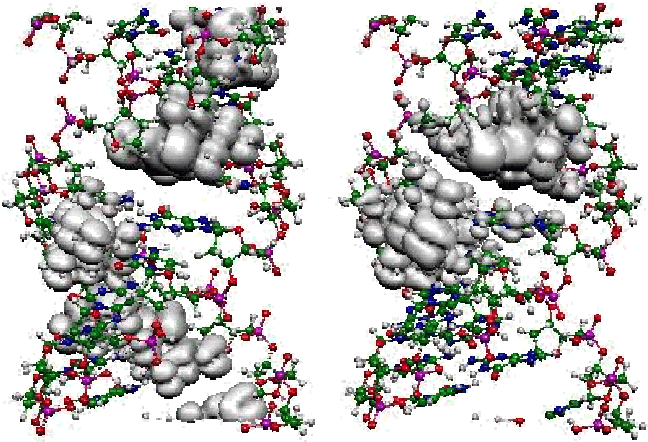}
\caption[]{Isosurface as in figure \ref{homo-lumo} for the top
eleven occupied states of the swapped structure (left). The same
for one particular state (the 11th) for a density value 50 times
smaller (right).} \label{homos}
\end{center}
\end{figure}

  The effect of the swap on the electronic structure of the chain
is dramatic, as can be seen in figure~\ref{bars}.
  The HOMO of the swapped guanine (indicated by an asterisk in
figure~\ref{bars}) sinks 0.55 eV (14 times the HOMO bandwith) into
lower valence band levels.
  This stabilization is of electrostatic origin ~\cite{ADNA-PRB}.
  Figure~\ref{homos} (left) shows the electronic density associated
to the eleven highest occupied states for the swapped structure,
showing the cut in the HOMO-state channel produced by the swapped
pair.
  The situation is similar for the unoccupied band, albeit less
dramatic, with the LUMO of the swapped cytosine raising 0.5 eV
above the LUMO band.

  All the results presented so far refer to fixed, relaxed
geometries.
  It has been proposed, however, that the electron transfer could
be mediated by phonons
\cite{Schuster99,Conwell00,Ladik99,Bruinsma00}.
  Indeed, there are many modes in DNA that are very soft and
related to the nucleic bases, since these flat, rigid molecules
hang from the backbone by a single bond.
  It is beyond the scope of this study to calculate phonon modes
and their interactions with electrons, but interesting qualitative
information on the matter can be inferred from the
conjugate-gradient (CG) minimizations.
  If close to the minimum, any snapshot during the minimization
gives the atomic positions that correspond to the freezing of
vibration modes with various amplitudes.
  The closer to the minimum, the more important the low-frequency
modes versus the high-frequency ones \cite{valley}.
  We study thus the electronic structure of geometries chosen from
the CG path with an energy of around 0.9 eV above the minimum,
that would correspond to room temperature or lower
\cite{temperature}.
  The highest occupied energy levels of such a geometry (called
ragged) is presented in figure~\ref{bars}.
  If not as dramatic as the swapping, the effect of these
``vibrations" on the electrons is also important, since the
guanine HOMO one-particle states spread over an energy width eight
times the original band width.

\section{Discussion and Conclusions}

  Electronic transport in DNA requires the presence of carriers.
  In controlled biochemical experiments (in solution), doping is
induced by intercalating or attaching to the DNA specific
molecules of known tendency to donate or accept electrons
\cite{Barton99}.
  In the dry experiments introduced earlier there is uncontrolled
doping of DNA.
  Since the negative charge of each phosphate group along both
chains has to be compensated by a cation from solution, in the
drying process cations precipitate along the DNA helix decorating
it.
  In addition to a very probable disorder in the cation positions
and variability in the amount of water molecules hydrating them,
there will also be deficiency or excess of counter-cations.
  Missing cations introduce holes in the valence band and
cations in excess (or cations with higher charge) introduce
electrons in the conduction band.
  The drying procedures of the dry experiments do not allow
estimating the concentration of carriers, not even their sign,
even if (in view of the ionization potential of G and the electron
affinity of C) the presence of holes has been argued to be more
likely.
  It is reasonable to expect in any case that carrier concentrations
will remain low, and that for the more drastic drying procedures
the likelihood is higher for the presence of holes, since cations
will be pulled away with the water.
  pGpC has been reported to be $p$-type in at least one
experimental setup \cite{Yoo01}.

  It has been shown above that the HOMO band width is one order of
magnitude smaller than the energy scale associated to sequence
disorder.
  In terms of a one-dimensional Anderson model, this leads
to electronic localization over very few base pairs.
  The localization is observed when plotting the separated
eigenstates.
  Figure~\ref{homos} (right) shows one electronic state within the
HOMO energy range.
  The isosurface plotted is for a density 50 times smaller that
that on the left, showing clearly the localization of the state
within a few base pairs.
  Sequence disorder will therefore dramatically depress the possible
band-structure-like DC conduction in DNA of arbitrary sequence
($\lambda$-DNA).
  The possibility \cite{Ratner00} of electron hopping or
tunneling, like in proteins \cite{Beratan97} is open, but it is
not expected to show the conductances close to one quantum
reported for $\lambda$-DNA by some groups \cite{Fink99,Kasumov01}.
  For electron doping, disorder fluctuations are still comparable
or larger than the LUMO band width, and the localization of the
electronic states is still considerable, albeit not as extreme as
for the hole-doped case.

  It is known \cite{Phillips91} that certain types of
disorder do not localize electrons as in the Anderson model.
  They represent, however, very particular realizations that
cannot be expected from $\lambda$-DNA in general.
  Very interesting recent results \cite{Carpena02} show possible
electron delocalization in 1D systems with more general long-range
correlated disorder, as the disorder of natural DNA.
  More research is needed to see how robust is this result when
including other effects like off-diagonal disorder or vibrations.
  In any case, the very large disorder-to-bandwidth ratio presented
above makes extended states for holes very unlikely, the typical
state being localized as displayed in figure~\ref{homos} (right).

  The particular experiment of Porath et al. \cite{Porath00} used
a controlled DNA sequence, namely pGpC, for which the sequence
disorder arguments do not apply.
  They explain their results in terms of band-like conduction.
  The results and analysis of Berlin and coworkers
\cite{Ratner00} propose band-like conduction in this particular
system as well.
  It is tempting to support this band-like condution assuming it
proceeds via electrons in the LUMO, given the agreement of the
widths of our LUMO band and the feature they observe
spectroscopically (the sensitivity of the LUMO to the environment
mentioned above has to be kept in mind, however).
  Such an explanation would be by no means conclusive in the light
of our ``vibration" results, since a deterioration of conduction
should be expected with temperature, which does not seem to
happen.
  The spread of HOMO states with frozen-in vibrations indicates
the energy scale of electron-phonon interactions, again very
substantial as compared with the band widths.
  Even if a more rigorous study is needed to quantify these
interactions and their effect, our rough estimations above allow
us to expect a substantial reduction of conduction with
temperature even below room temperature.

  Considering the floppy modes in DNA and their effect on
the electron states, and the fact that a hole on a guanine
represents a positive charge amidst negatively charged guanines,
the idea of self trapped (small) polarons seems not at all unlikely.
  Polarons have already been suggested \cite{Schuster99,Conwell00}
and supported \cite{Yoo01} in the literature.
  Some of the models used have assumed the polarons to be of an
origin similar to that found in polyacetylene, i.e. of
off-diagonal (hopping) origin.
  Polarons of electrostatic origin are described by an electronic
Hamiltonian in which the diagonal matrix elements are the ones
that change with inter-base separation and stabilize the polaron,
as in Holstein's model \cite{Holstein59}.
  A purely classical electrostatic model based on the ab initio
results for the geometry and ground-state electrostatic
characteristics of the system \cite{ADNA-PRB} displays polaronic
charge excitations with a spread of a few base pairs.
  DFT calculations on the matter will be presented elsewhere
\cite{Simone}.

\vspace{10pt}

{\it Acknowledgments.}
  We are grateful to J. $\rm{\check{S}poner}$ and R.
Santamar{\'{\i}}a for making their coordinates of bases and
base-pairs available to us.
  We are also indebted to R. Weht for many useful discussions and
his help during the first stages of this work.
  This work has been supported by the Spanish Ministerio de Ciencia
y Tecnolog\'{\i}a under grant BFM2000-1312, and by the
Fundaci\'on Ram\'on Areces.
  DSP acknowledges support from Spain's MCyT under the Ram\'on y
Cajal program.
  Calculations were performed at the CCCFC of the Universidad
Aut\'onoma de Madrid and at the PSMN of the \'Ecole Normale de
Lyon (SESC).


\begin{references}
\bibitem{radii}
The cutoff radii for the finite range atomic orbitals used in this
work are (in a.u.): 4.2 for H, 4.1 for C, 3.6 for N, 3.2 for O,
and 4.7 for P. For the atoms intervening in the hydrogen bonds
longer radii were used (in a.u.): 5.5 for H, 4.6 for N, and 4.2
for O. The shapes of the basis functions and the radii of the
double-$\zeta$ orbitals are as in \cite{Emilio99} with a
split-norm parameter of 0.15.
  The possibility that the finite range of the basis orbitals could be
influencing the results has been tested by recalculating the
electronic structure of the ordered structure using orbital ranges
increased by 30\%. The results did not change significantly (bandgap
2.14 eV, HOMO bandwidth 30 meV, LUMO bandwith 265 meV).
\bibitem{temperature}
Assuming a canonical energy distribution among the low-frequency
modes only (3 per nucleic basis).
 If the energy is distributed among more modes than the
ones assumed the temperature is lower.
\end{references}
\end{document}